\pdfoutput=1
\documentclass[prd,aps,showpacs,preprintnumbers,amsmath,amssymb,nofootinbib]{revtex4}
\usepackage{graphicx}
\usepackage{epsf}
\usepackage{amsmath}
\usepackage{amssymb}

\usepackage{graphicx}
\usepackage{dcolumn}
\usepackage{bm}
\def\be{\begin{equation}}
\def\ee{\end{equation}}
\def\bea{\begin{eqnarray}}
\def\eea{\end{eqnarray}}

\begin{document} 

\title{Multi-Stream Inflation in a Landscape}

\author
{
  Francis Duplessis \thanks{francis.duplessis@mail.mcgill.ca},
  Yi Wang \thanks {wangyi@hep.physics.mcgill.ca},
  Robert Brandenberger \thanks{rhb@hep.physics.mcgill.ca}
  \\
  \textit{Physics Department, McGill University, Montreal, H3A2T8,
    Canada}
}


\begin{abstract}

There are hidden observables for inflation, such as features localized in position space, which do not 
manifest themselves when only one inflation trajectory is considered. To address this issue, we 
investigate inflation dynamics in a landscape mimicked by a random potential. We calculate the 
probability for bifurcation of the inflation trajectory in multi-stream inflation. Depending on the shape of the
random bumps and the distance between bumps in the potential, there is a phase transition: on one 
side of the critical curve in parameter space isocurvature fluctuation are exponentially amplified and 
bifurcation becomes very probable. On the other side bifurcation is dominated by a random walk 
where bifurcations are less likely to happen.

\end{abstract}

\maketitle

\section{Introduction}

During multi-stream inflation \cite{Li:2009sp, Li:2009me, Wang:2010rs, Afshordi:2010wn}, the classical 
inflation trajectory bifurcates into multiple branches. Multi-stream inflation provides a possible 
explanation for features and asymmetries in the CMB, e.g. non-Gaussianities \cite{Li:2009sp, Wang:2010rs}, 
cold spots in the CMB and great voids in the large scale structure \cite{Afshordi:2010wn}, and also have a 
web implication for eternal inflation \cite{Li:2009me}.

Although multi-stream inflation is not necessarily embedded into string theory, the motivation is best 
understood in terms of the string landscape 
\cite{Bousso:2000xa, Giddings:2001yu, Kachru:2003aw, Douglas:2003um}. According to the string 
landscape paradigm, string theory is unique while the vacuum structure of string theory is extremely 
complicated. Based on semi-classical investigations of string compactifications one estimates that
there are of order $10^{100}$ or more meta-stable vacua. This forms the string landscape. The 
implications of the string landscape for inflation should be considered seriously.

Over the past decade there has been a lot of work on inflation in the context of string compactifications
(see e.g. \cite{Dvali, Stephon, Burgess} for some early papers and e.g \cite{stringinflationrevs} for
some recent review articles). A lot of the work has focused on specific string theoretical constructions
which might yield inflation, and has not taken the landscape features of string theory into account.
However, there have been a number of inflation scenarios motivated specifically by the string landscape. 
An incomplete list includes 
inflation in a random potential \cite{Tye:2008ef, Agarwal:2011wm, Frazer:2011br},
chain inflation \cite{Freese:2004vs, Huang:2007ek, Chialva:2008zw, Ashoorioon, Cline:2011fi}, 
particle production during inflation \cite{Chung:1999ve, Romano:2008rr, Barnaby:2009dd, Barnaby:2010ke}, 
extra symmetry point encounter \cite{Battefeld:2010sw, Battefeld:2011yj}, meandering 
inflation \cite{Tye:2009ff}, multi-stream inflation and 
quasi-single field inflation \cite{Chen:2009we, Chen:2009zp}.

In this paper, we focus on multi-stream inflation. It was argued that in the string landscape, where the 
inflation potential is extremely complicated, bifurcations in the inflation trajectory become possible or 
even frequent. But no quantitative justifications for this statement were previously provided. The aim 
of the present paper is to fill this gap, and bring a connection between the landscape and observation.

The observational consequences fall into two possibilities: When the bifurcation probability is rather 
small, there may be only a few bubbles in the sky following exotic trajectories. In this case multi-stream 
inflation provides a possible explanation of the cold spot on the CMB, or a void in the large scale structure. 
On the other hand, when the bifurcation probability is large, extra fluctuations are not under control and 
thus such random potentials are not consistent with observations -- except when they are extremely 
finely tuned \footnote{Here we are having in mind a random potential. If the bifurcation potential is 
not random (say, originate from spontaneous breaking of an approximate symmetry), 
common bifurcations do not necessarily require a fine tuning.}, or when the time of bifurcation is pushed to before the observable period of inflation (the
final period of inflation during which scales which are currently observed exit the Hubble radius). 

The inconsistency for high probability bifurcations in a random potential comes in at least two aspects: Most importantly, different trajectories typically have different e-folding numbers. the e-folding number difference between different trajectories results in an extra fluctuation in the scalar type perturbations. Such a fluctuation with amplitude much greater than $10^{-5}$ is already ruled out on the scales of CMB. Also, if the different trajectories do not combine or reheat into the same radiation, there will be domain walls in between them, which causes a problem. 

In the string landscape the moduli space of low energy modes may have a very large dimension. For 
simplicity, we assume in this paper that only two fields participate in the dynamics during the
inflationary phase. All other flat directions are taken to be stabilized. They do not evolve dynamically 
but could provide bumps in the effective two dimensional landscape. If there were more flat 
directions taking place in the inflationary dynamics, the bifurcation probability should be larger. 
Thus, in our present analysis we are actually under-estimating what the bifurcation probability would be
in a more general case, perhaps by a huge factor.

The paper is organized as follows: In Section 2, we model the landscape and calculate the bifurcation 
probability in the landscape analytically. Two cases are considered, either when bifurcations 
are dominated by an exponential growth of isocurvature fluctuation, or when they are given 
by a random walk of the isocurvature field. In Section 3, we use stochastic equations to numerically 
verify the analytical approximations made in the previous section. We discuss our results in Section 4.

\section{Bifurcation Probabilities in a Landscape}

In this section, we calculate bifurcation probabilities in the landscape. We show that there is a 
phase transition controlled by the height and the ellipticity of the bumps in the potential.

To model the random potential, we focus on a double-field model. The fields are labeled 
$\varphi_1$ and $\varphi_2$. There is a overall slope in the $\varphi_1$ direction, such that 
if we remove the randomness of the potential, $\varphi_1$ becomes the inflation direction and 
$\varphi_2$ has flat potential. To add random features to the potential, we 
introduce a perturbation $U$ of the potential which is characterized by randomly 
located and separated bumps. These bumps are characterized by field values 
$\Delta_p \varphi_1$ and $\Delta_p \varphi_2$ which denote the mean size 
and separation of the bumps in the 
$\varphi_1$ and $\varphi_2$ directions respectively. An ellipticity parameter 
$\xi \equiv \Delta_p \varphi_1 / \Delta_p \varphi_2$ is defined to model the shape of the bumps.
The amplitude of the bumps is taken to be the limiting value such that the
slow roll condition in the inflaton direction is maintained.

The equations of motion for the fields take the form
\begin{align}
  &
  \ddot \varphi_1 + 3H \dot\varphi_1 + \partial_1 V(\varphi_1) + \partial_1 U(\varphi_1,\varphi_2) =
  0~,
  \\
  &
  \ddot \varphi_2 + 3H \dot\varphi_2 + \partial_2
  U(\varphi_1,\varphi_2) = 0~,
  \label{eq:phi2}
\end{align}
where $V(\varphi_1)$ is the slow roll potential, subject to slow roll restrictions, and 
$U(\varphi_1,\varphi_2)$ is a random potential. Shorthand notations $\partial_1$ and 
$\partial_2$ are used to denote the partial derivatives with respect to the two fields. 
We use a parameter $\lambda$ to denote the ratio 
\begin{equation}
\lambda \equiv \sqrt{\langle (\partial_1U)^2\rangle}/|\partial_1V| \, .
\end{equation}
When $\lambda\ll 1$, the slow roll condition in the inflation direction remains satisfied.

\subsection{Definition of a bifurcation}

One subtlety is the question of quantifying what is a bifurcation. We adopt the following two 
definitions: a bifurcation happens when two sample inflation trajectories (with the same initial 
conditions):
\begin{enumerate}
\item{End on different sides of a bump. }\label{option1}
\item{Are much farther apart at the end of the inflationary phase than would be expected 
from quantum fluctuations alone. } \label{option2}
\end{enumerate}
Numerically, we find that the bifurcation probabilities obtained from these two definitions are practically equal in the interesting region of parameter space. The two definitions give different results when the size of the bumps (both in terms of field extent and height) becomes too small to have any impact on the dynamics. In that case the probability drops to zero for definition \ref{option2} while can stay high for definition 1. As we will see, the size when the bumps in the potential become negligible is when they become smaller than the expected quantum fluctuations of the fields.  With that insight we define bifurcations based on Definition \ref{option2} \footnote{Definitions 1 and 2 also differ when the field extent of the bump is as large as the field distance which the inflaton field moves in of order of a few e-foldings. However, in this case we cannot speak of a random potential any more inasfar as the dynamics relevant to cosmological observations is to be considered. Thus, here we are not interested in this situation.}.

\subsection{Bifurcation from instabilities}
\label{sec_instabilities}

Consider two trajectories of $\varphi_2$, denoted by $\varphi_2^{(A)}$ and $\varphi_2^{(B)}$. 
Let the deviation of these two trajectories as $\delta \equiv \varphi_2^{(A)} - \varphi_2^{(B)}$. 
Then as long as $\delta$ is smaller than $\Delta_p\varphi_2$, one can expand $U$ such that 
$\delta$ satisfies
\begin{equation} \label{deveq}
  \ddot \delta + 3 H \dot\delta + (\partial_2^2 U) \delta = 0 ~.
\end{equation}

Now we discuss general features of the potential $U$. It follows from the classical
equation of motion and from the normalization of the density fluctuations produced
during inflation that during one e-folding of inflation $\varphi_1$ direction rolls 
a distance of order $10^5 H$. \footnote{This is because, the rolling distance of $\varphi_1$ per Hubble time is $\dot\varphi / H \sim \left( H^2/\dot\varphi \right)^{-1} \times H \sim P_\zeta^{-1/2} H$.}
This distance is extremely long from the point of
view of the field $\varphi_2$, because of the smallness and randomness of the 
potential $U$. Thus it is a good approximation to describe the dynamics of $\varphi_2$
as taking place in a time-dependent potential given by
\begin{equation} \label{Upot}
  U(\varphi_1,\varphi_2) = U(\varphi_1(0) + \dot \varphi_1 t, \varphi_2)~,
\end{equation}
where $\varphi_1(0)$ is the initial value of the field and $\dot\varphi_1$ can be treated 
as a constant. 

Over a time interval of a few Hubble times, the field trajectory can pass through a lot of 
bumps. The number of bumps the field trajectory crosses during a time interval $t$ is 
about $\dot\varphi_1 t / \Delta_p\varphi_1$. Thus, on time scales of order 
$\Delta_p\varphi_1 / \dot\varphi$, the $\partial_2^2$ parameter in equation 
(\ref{deveq}) will typically change its sign. To model this behavior, we can use
the following approximation for the form of the second derivative of the potential
$U(\varphi_2, t)$ from (\ref{Upot}):
\begin{equation}\label{eq:p2uapprox}
  \partial_2^2 U \simeq \frac{\lambda\xi \partial_1 V}{\Delta_p\varphi_2}
  \sin\left(\frac{2\pi \dot\varphi_1 t}{\Delta_p\varphi_1}\right)~.
\end{equation}
When the sine function change its sign, the equation (\ref{deveq}) changes its behavior 
from oscillation to exponentially growth. Now the question is whether the overall behavior 
shows exponential growth or not. Qualitatively, we see that when the sine function changes 
its sign slowly enough, there is enough time for growth to take place. On the other hand, when 
the sign function change its sign extremely quickly, the equation (\ref{deveq}) can be treated 
adiabatically thus there should be no growth. 

\begin{figure}[htbp]
\centering
\includegraphics[height=0.22\textheight]{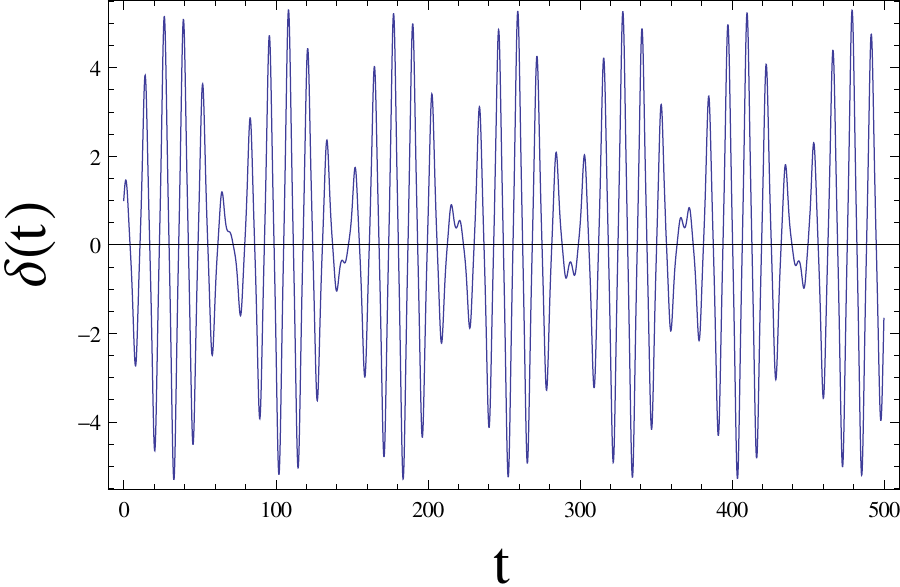}
\includegraphics[height=0.22\textheight]{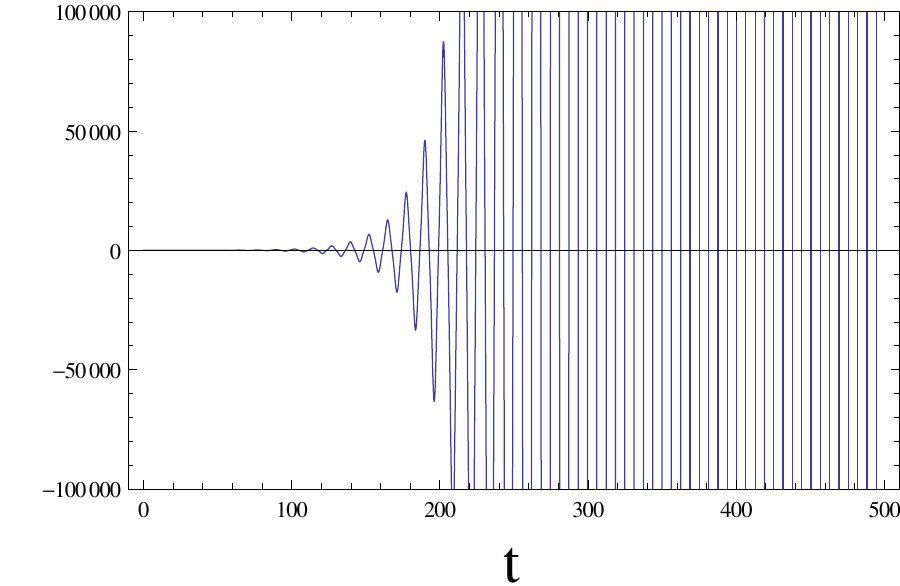}
\caption{\label{fig:4546} The behavior of equation $\ddot \delta +
  \alpha \sin(\beta t) \delta = 0$. Here $\beta=1$ is chosen without
  loss of generality because varying $\beta$ is a rescaling of time. The
  left panel has $\alpha/\beta^2=0.45$, and the right panel has
  $\alpha/\beta^2=0.46$. A sharp transition is observed: for $\alpha=0.45$,
  the amplitude of $\delta$ does not grow, while for $\alpha=0.46$,
  the amplitude grows exponentially. }
\end{figure}

To be a bit more precise, if we neglect the 
Hubble friction term for a moment (we will return to the full treatment at
the end of this subsection), the equation (\ref{deveq}) becomes
\begin{equation}
\ddot \delta + \alpha \sin(\beta t) \delta = 0 
\end{equation}
and thus has the form of the Mathieu equation, the same equation
which describes preheating after inflation (see \cite{TB, KLS} and
\cite{Allahverdi} for a recent review). It will have exponentially growing 
solutions when 
\begin{equation} \label{cond}
|\alpha| / \beta^2 > 0.45
\end{equation}
(as shown in figure \ref{fig:4546}), and non-growing solution otherwise. 
Note that (\ref{cond}) is the condition for the time scale of the change
of the sign of the frequency to be longer than the intrinsic time scale of
oscillation.

Inserting the parameters describing the size of the
random potential, the condition (\ref{cond})
to have growing solution in the $\varphi_2$ direction becomes
\begin{equation}
  \left|\frac{\lambda\xi \partial_1 V}{\Delta_p\varphi_2}\right|
  > 0.45 \left(\frac{2\pi \dot\varphi_1}{\Delta_p\varphi_1}\right)^2 ~.
\end{equation}
Note that the slow roll condition in the inflaton direction implies
$|\partial_1 V| \simeq |3H\dot\varphi_1|$. Making use of the
result $P_\zeta^{1/2} = H^2 /(2\pi\dot\varphi)$
for the power spectrum of cosmological perturbations $\zeta$ to
eliminate ${\dot{\varphi_1}}$, the above inequality becomes
\begin{equation}
  \frac{\Delta_p\varphi_1}{H} > \frac{1}{\lambda \xi^2 P_\zeta^{1/2}}~.
\end{equation}
When this inequality is satisfied, the solution turns out to be exponentially growing. 
In order for the exponential growth rate to be large on compared to
the Hubble expansion rate we need
\begin{equation}
  \left|\frac{\lambda\xi \partial_1 V}{\Delta_p\varphi_2}\right| > H^2~.
\end{equation}
This requirement can be rewritten as
\begin{equation}
  \frac{\Delta_p\varphi_1}{H}<\frac{3\lambda\xi^2}{2\pi P_\zeta^{1/2}}~.
\end{equation}
Thus when
\begin{equation}\label{eq:bifrange}
  \frac{1}{\lambda \xi^2 P_\zeta^{1/2}} <
  \frac{\Delta_p\varphi_1}{H}<\frac{3\lambda\xi^2}{2\pi
    P_\zeta^{1/2}}~,
\end{equation}
a perturbation in the isocurvature direction will grow until its size reaches the
value $\Delta_p\varphi_2$ when bifurcations become common. Note that in 
terms of $\xi$, both the inequalities give lower bounds.

Now consider the full equation (\ref{deveq}) including the Hubble friction term. 
Experience from studies of preheating (see \cite{Allahverdi} for a recent review)
teach us that an instability derived in the absence of Hubble friction persists when
restoring the friction term: via a field rescaling with a power of the cosmological
scale factor the friction term can be eliminated, and the equation of motion of
the rescaled variable remains of the type described by Floquet theory and
showing an instability. 

\begin{figure}[htbp]
\centering
\includegraphics[width=0.8\textwidth]{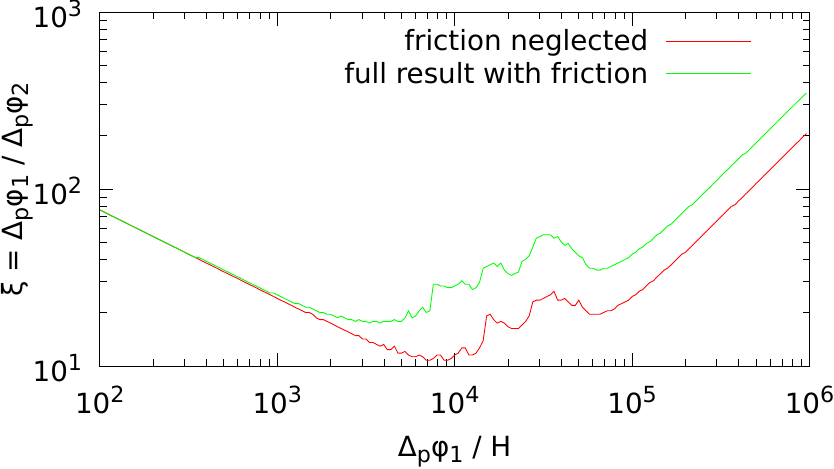}
\caption{\label{fig:comp} Comparison between the results neglecting the
  $3H\dot\varphi_2$ term (lower curve) and the full result (upper curve).  
  Here the value $\lambda=0.1$ is
 used. The estimate neglecting friction overestimates the bifurcation
  probability by a factor of up to 1.7 in the variable $\xi$.}
\end{figure}

\begin{figure}[htbp]
\centering
\includegraphics[width=0.8\textwidth]{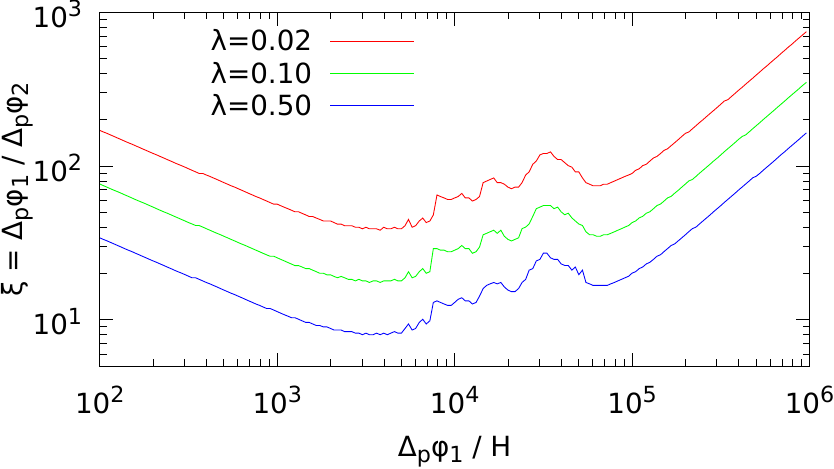}
\caption{\label{fig:exp} The lower bound on $\xi$ in order to have
  bifurcation of the amplification type. Values of $\xi$
  above the red, green, blue lines (top line is the red line, bottom line is
  the blue line) correspond to regions in
  parameter space where exponential amplification occurs
  for values of $\lambda=$0.02, 0.10, 0.50 respectively. In these
  regions of amplification, the bifurcation probability becomes of order one.}
\end{figure}

If there is to be an observable bifurcation, we require that after about 
10 e-foldings the growth 
factor times  $H/(2\pi)$ (which is the typical magnitude of quantum fluctuations in a
de Sitter phase and which we take to be the initial value of $\delta$) becomes greater 
than the separation $\Delta_q \varphi_2$ between bumps in the random potential $U$ . 
This condition can be analyzed numerically, and the result is shown in Figure \ref{fig:comp} 
and Figure \ref{fig:exp}, as a lower bound on $\xi$ to get bifurcation leading to amplification
of the fluctuations. In both figures the area above the curve yields the parameter space
for which bifurcations occur and have an observable effect.

From Figure \ref{fig:comp}, we observe that for small $\Delta_p\varphi_1 / H$, one can 
safely neglect $3H\dot\varphi_2$. This is because in that case the oscillations have a
much higher frequency than Hubble friction. On the other hand, when 
$\Delta_p\varphi_1 / H$ is large, neglecting $3H\dot\varphi_2$ results in an over-estimation 
of $\xi$ by a factor of $1.7$. Figure \ref{fig:exp} compares the lower bound for bifurcations 
obtained using different values of $\lambda$.

\subsection{Correction factors}

In the previous subsection, we discussed the basic picture of amplified bifurcations 
in a random potential. However, the model we used is over-simplified, in the sense 
that there are factors of order one (or order 10) which were ignored. Since we want 
to relate our analytical analysis to numerical simulations we must discuss the
various correction factors

First, as will be discussed in Section \ref{sec_randompotential}, the numerical simulations 
are performed with the maximal amplitude $A$ of the perturbation $U$ of the potential
being a function of the bump size, i.e. $A(\Delta_p\varphi_1,\Delta_p\varphi_2)$ (chosen
such that the motion in the $\varphi_1$ direction is at the borderline of slow rolling). 
In the analytic calculation this corresponds to working with $\lambda = 1/(10\xi)$.

Second, there is a correction factor along a single inflation trajectory. We have used 
$\Delta_p\varphi_1$ and $\Delta_p\varphi_2$ to denote the average distance between 
bumps in the $\varphi_1$ and $\varphi_2$ directions. However, the criterion for amplified 
bifurcation is actually not controlled by the average distance, but rather the longest distance 
between bumps of the same sign in the potential.

To put it more explicitly, along an inflation trajectory, we have assumed that the derivative 
of the potential in the isocurvature direction $\partial_2 U$ changes its sign when the 
inflaton rolls an average distance $\Delta_p\varphi_1$. However, a random potential 
is not exactly periodic. As a result, on a distance of order $n \Delta_p\varphi_1$
(where $n$ is a positive integer), 
the sign of $\partial_2 U$ as a function of time will be a sequence with 
length $n$, like $(+,+,-,+,-,-,-,-,+,\cdots)$. The probability of bifurcation is determined by 
the longest sub-sequence without a sign change (4 in the above example). The determination
of this critical distance is a well defined mathematics problem. Here we do not give an 
analytical solution. However, numerically, we find that the expectation value 
(as an ensemble average) of the longest sub-sequence without sign change 
$\langle n_\mathrm{sub} \rangle$ can be approximated by
\begin{equation}\label{eq:contineous}
  \langle n_\mathrm{sub} \rangle = 1.5 \log(n)~, \qquad (n \gg 1)~.
\end{equation}
If we are interested in $\Delta N$ e-foldings of inflation, $n$ can be written as
\begin{equation}
  n = \frac{\dot\varphi \Delta N}{H \Delta_p\varphi_1} =
  \frac{\Delta N}{2\pi \sqrt{P_\zeta}} \frac{H}{\Delta_p\varphi_1}~.
\end{equation}
For example, if we take $\Delta N = 10$, use the CMB normalization of the curvature
power spectrum, and take $\Delta_p\varphi_1 /H= 100$, 
we have $n=324$ and thus $\langle n_\mathrm{sub} \rangle = 8.7$. For $\Delta_p\varphi_1 /H= 10^3$, 
we have $\langle n_\mathrm{sub} \rangle = 5.2$.

\begin{figure}[htbp]
\centering
\includegraphics[width=0.7\textwidth]{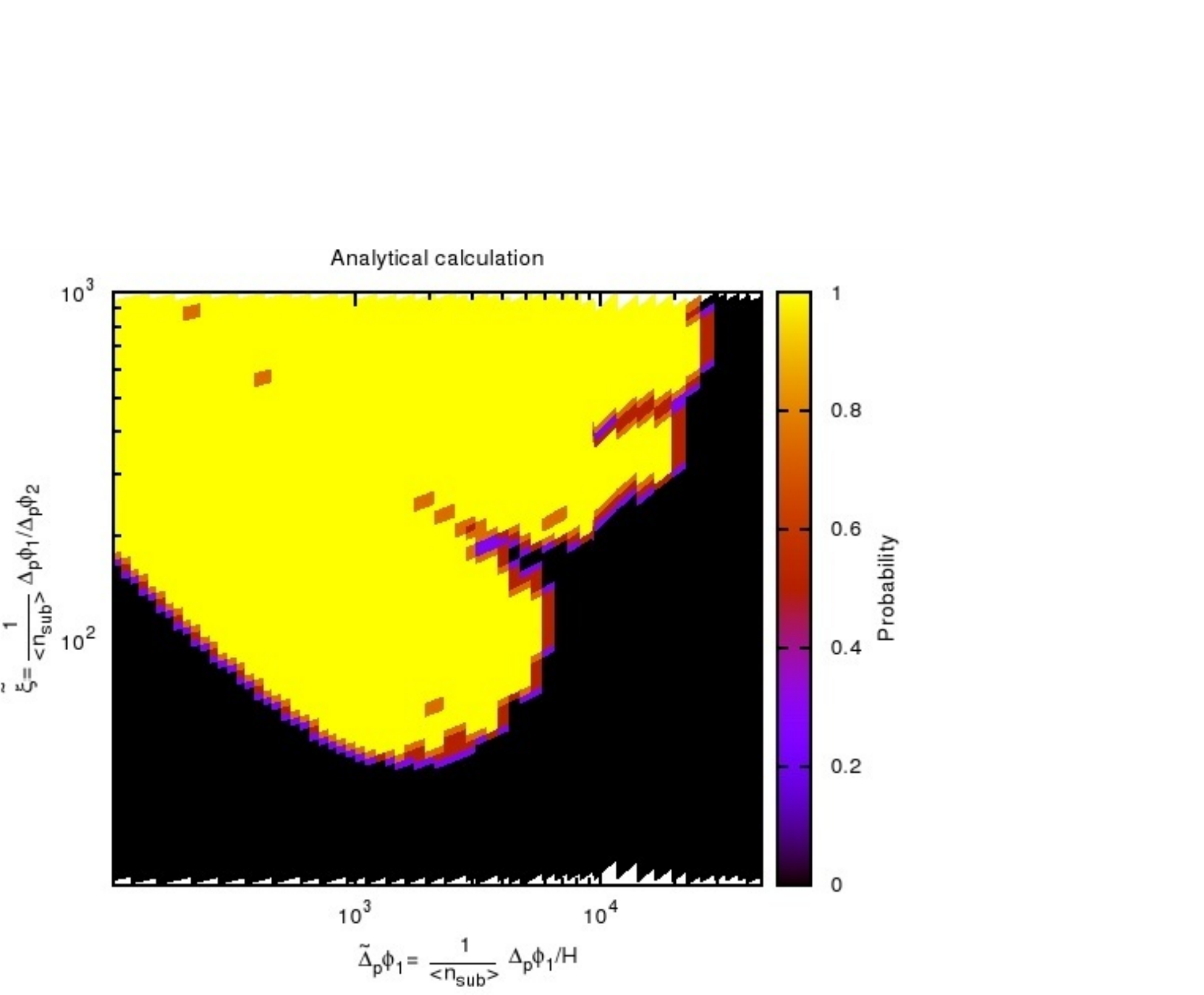}
\caption{Probability of bifurcation from analytical calculation simulations. We have 
rescaled data according to equation (\ref{eq:rescale}). Also, a color plot is made 
for the purpose of comparison with numerical results, with the same set of 
parameters.}\label{fig_heatmap_a}
\end{figure}

This means that if $\Delta_p\varphi_1$ is the mean size of the bumps in a random potential, 
then we should compare our numerical simulations to analytic results with bump taken to 
have size $\langle n_\mathrm{sub} \rangle \Delta_p \varphi_1$. Alternatively, if 
$\Delta_p\varphi_1$ is the size of the bumps in the analytic analysis, we must then 
compare the analytical results to simulations in a random potential with bumps of mean 
size $1/ \langle n_\mathrm{sub} \rangle \Delta_p\varphi_1$. This 
$\langle n_\mathrm{sub} \rangle$ factor yields a correction of the parameters 
$\Delta_p\varphi_1$, $\xi$ of the form:
\begin{eqnarray}\label{eq:rescale}
 \Delta_p\varphi_1 &\quad \rightarrow\quad  \tilde{\Delta}_p\varphi_1 \equiv  \frac{1}{ \langle n_\mathrm{sub} \rangle} \Delta_p\varphi_1 \\
 \xi \equiv \frac{\Delta_p\varphi_1}{\Delta_p\varphi_2} &\quad\rightarrow\quad \tilde\xi \equiv \frac{1}{ \langle n_\mathrm{sub} \rangle} \frac{\Delta_p\varphi_1} {\Delta_p\varphi_2}~,
\end{eqnarray}
and should be used in equations (\ref{eq:p2uapprox}) $\sim$ (\ref{eq:bifrange}).  The 
bifurcation probability using these first two correction factors is plotted in 
Figure \ref{fig_heatmap_a}. For large $\langle n_\mathrm{sub} \rangle$, we have used 
the above approximated formula (\ref{eq:contineous}), and for small 
$\langle n_\mathrm{sub} \rangle$, we have used the explicit result.

Third, in the previous subsection we were comparing only two trajectories. However, 
there are exponentially many trajectories (local Hubble volumes) created during inflation. 
Thus there may be rare bifurcations in some exceptional trajectories. To take those 
exceptional trajectories into account, note that the quantum fluctuations are nearly 
Gaussian and thus given by the following probability distribution
\begin{equation}
  \mathrm{Prob}\propto \exp(-\Delta\varphi^2/2\sigma^2)~,
\end{equation}
where $\sigma$ is the variance of the fluctuations. On the other hand, there 
are $\exp(3\Delta N)$ trajectories, where $\Delta N$ is the e-folding number 
difference between the bifurcation point and the start of observable inflation. 
Thus, typically, the largest fluctuation among these $\exp(3\Delta N)$ trajectories 
is $\Delta\varphi \sim \sigma  \sqrt{3\Delta N}$. For example, if we consider 
$\Delta N = 10$, where bifurcation takes place on the smallest observable scales 
on CMB, the rare bifurcations could take place even when $\Delta_p\varphi_1$ and 
$\Delta_p\varphi_2$ are 5 times greater than their critical values.

As mentioned in the introduction, there are big observational differences between the case of rare bifurcations and that of common bifurcations. The case of rare bifurcations could be consistent with current CMB observations and could in fact be responsible for a position space non-Gaussianity. However, if bifurcations become common, fine tuning is needed to make the scenario consistent with observations. This is because in a random potential, different trajectories commonly lead to different number of e-folds of inflation. The difference in e-folding number is translated into temperature anisotropies in the CMB via the $\delta N$ formalism, which will not be consistent with observations when $\delta N \gg 10^{-5}$.

\subsection{Bifurcation from a Random Walk}

Now let us proceed to consider the case in which bifurcations do not lead to an amplification of
the isocurvature fluctuations. In this case, we define $\Delta t$ as the time duration for the 
field to cross the bump in $\varphi_1$ direction. Thus 
\begin{equation}
\Delta t = \Delta_p\varphi_1 / \dot\varphi \, .
\end{equation}
During the time interval $\Delta t$, the quantum fluctuation in the $\varphi_2$ direction is
\begin{equation}
  \Delta_q\varphi_2 = \frac{H}{2\pi} \sqrt{H \Delta t}~.
\end{equation}
The bifurcation probability during this $\Delta t$ time is determined by the ratio
\begin{equation}\label{eq:bifur-Dt}
  \textrm{Prob}(H\Delta t) = \frac{\Delta_q\varphi_2}{\Delta_p\varphi_2}
\end{equation}
when $\Delta_q\varphi_2 < \Delta_p \varphi_2$. Otherwise, when 
$\Delta_q\varphi_2 > \Delta_p \varphi_2$, bifurcations happen frequently, but on the 
other hand the bumps are too small to lead to interesting consequences. 

Now, consider a world line of a comoving observer. The probability for the observer 
to see bifurcation(s) during $N_e$ e-foldings is
\begin{equation}
  \textrm{Prob}(N_e) = \sum_{n=1}^{N_e/(H\Delta t)} [1-\sqrt{n}\textrm{Prob}(H\Delta t)]~.
\end{equation}
The $\sqrt{n}$ factor comes from the random walk accumulation. To have a considerable 
bifurcation probability, we need
\begin{equation}
  \frac{2}{3}\left(\frac{N_e}{H\Delta t}\right)^{3/2}
  \mathrm{Prob}(H\Delta t) \geq 1~.
\end{equation}
Inserting numbers, we obtain the following condition for bifurcation
\begin{equation}
  \frac{\Delta_p\varphi_1}{H}\leq \sqrt{\frac{N_e^{3/2}\xi}{6\pi^2
      P_\zeta^{1/2}}}~.
\end{equation}
For 10 e-foldings, which corresponds to observable scales on the CMB, and for $\xi=1$, 
the above inequality is saturated for the value $\Delta_p\varphi_1 / H \simeq 100$.

In this subsection, we have not considered the classical dynamics of the $\varphi_2$ field. 
Considering the $\varphi_2$ field may encounter a bump and bounce back, we 
may have over-estimated the bifurcation probability.
Moreover, if we impose the slow-roll constraint $\epsilon\ll 1$, bumps formed of Lagrange 
polynomials will have a negligible effect on the fields compared to quantum fluctuations 
when $\Delta_q\varphi_2 / \Delta_p\varphi_2 \sim 1$. Hence we do not expect any bifurcations 
if we stick with the definition \ref{option2}. The conclusion of the current subsection is that, 
without amplification, bifurcation is less probable to happen in a way which has
observational consequences.

\section{Simulation for Stochastic Inflation}

Numerical simulations were performed to obtain the probability of bifurcation in 
random potentials. The dynamics of our fields are created using Starobinsky's 
stochastic approach \cite{Starob} to take into account the quantum fluctuations. 
In this approach, the equations of motions for $\varphi_1$ and $\varphi_2$ take the form,
\begin{equation}\label{eq_stochastic1}
\ddot\varphi_1(t)+3 H \dot\varphi_1 (t) + V_{,\varphi_1}=\frac{3}{2\pi} H^{5/2}\eta_1(t),
\end{equation}
\begin{equation}\label{eq_stochastic2}
\ddot\varphi_2(t)+3 H \dot\varphi_2 (t) + V_{,\varphi_2}=\frac{3}{2\pi} H^{5/2}\eta_2(t).
\end{equation}
Here the $\eta_i$ terms are stochastic sources which obey independent Gaussian 
distributions and are normalized to give
\begin{equation}\label{eq_sourcesnormalisation}
\langle \eta_i(t)\eta_i(t') \rangle=\delta(t-t') \, .
\end{equation}
To evolve the fields numerically we discretize time in intervals $\Delta t=t_n-t_{n-1}$ 
so that the delta function becomes
\begin{equation}\label{eq_discretedelta}
\delta(t-t')\rightarrow \frac{\delta_{nm}}{\Delta t}.
\end{equation}
We do not use the approximations $|\ddot\varphi_1| \ll |3 H \dot\varphi_1|$ during 
the simulation but we did make use of the approximation that the energy density 
is dominated by the potential in order to write $H=\sqrt{V/3}$.

\subsection{Random Potential} \label{sec_randompotential}
 
Our random potentials can be written as,
\begin{equation}
V(\varphi_1)+U(\varphi_1,\varphi_2)~,
\end{equation}
where $V$ is a background potential that has a small tilt in the $\varphi_1$ direction. We 
chose $V=\frac{1}{2}m^2 \varphi_1^2$ for the numerical work. The second term 
$U$ is a set of random perturbations which are created by discretizing field space 
$(\varphi_1,\varphi_2)$ into a lattice with separation\footnote{A given lattice spacing 
will represent a potential which is dominated by perturbations whose mean size is of the 
lattice spacing, hence the use of $\Delta_p\varphi_i$ as notation.} $\Delta_p\varphi_1$ and 
$\Delta_p\varphi_2$ and assigning a random number from the interval $[-A,A]$ at 
each point, and then performing a two dimensional interpolation to obtain a continuous function.

The value of $A(\Delta_p\varphi_1,\Delta_p\varphi_2)>0$ represents the amplitude of the 
perturbations and depends on the lattice spacing because we require that the full potential 
$V+U$ does not lead to violation of the slow-roll condition on the $\epsilon$ parameter. 
This constraint can be made explicit when the form of the perturbations are known. In our 
case the perturbations are interpolations between the random values associated to each 
lattice point using a second order Lagrange polynomials.

\begin{figure}[htbp]
\centering
\includegraphics[width=0.45\textwidth]{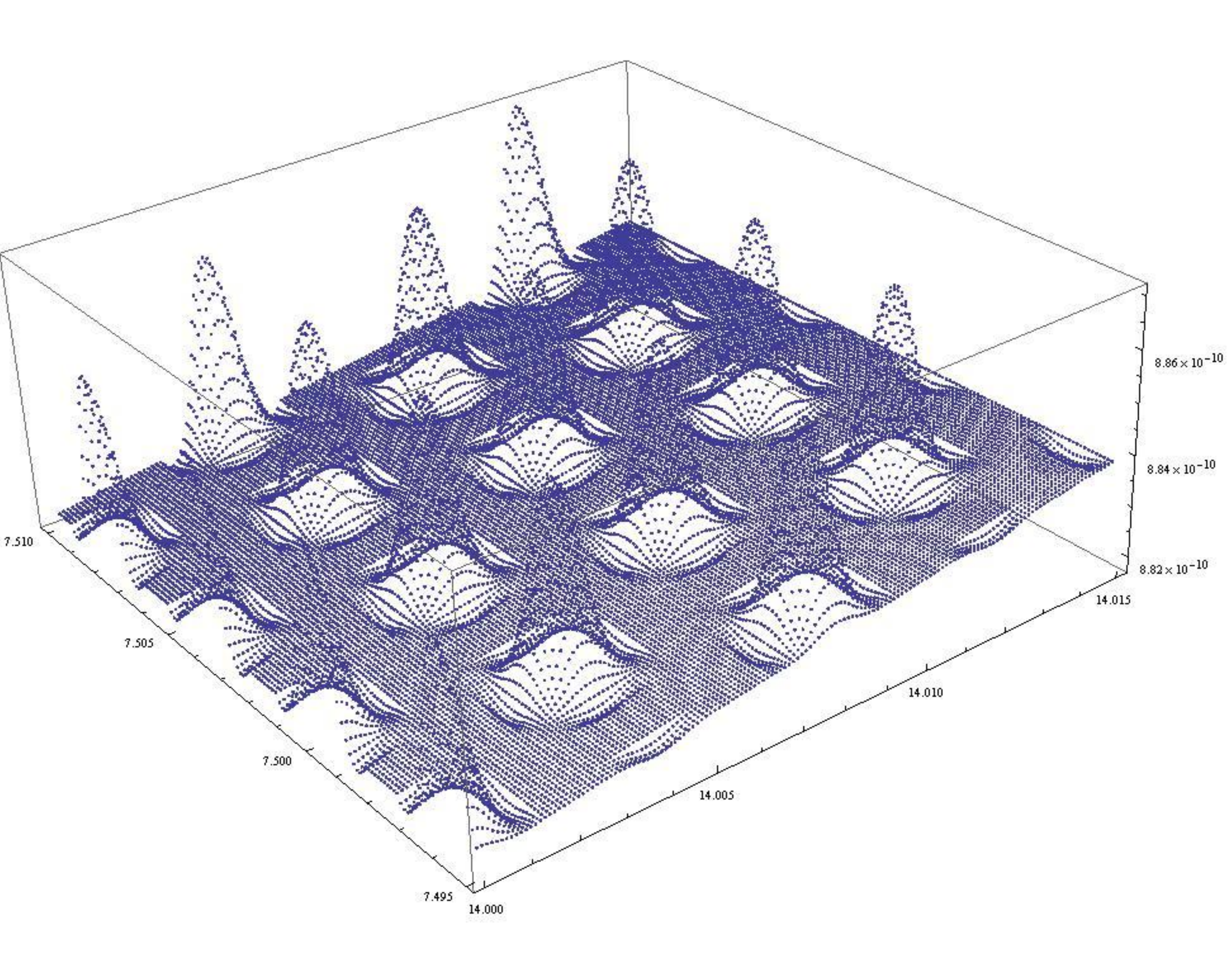}
\hspace{0.05\textwidth}
\includegraphics[width=0.45\textwidth]{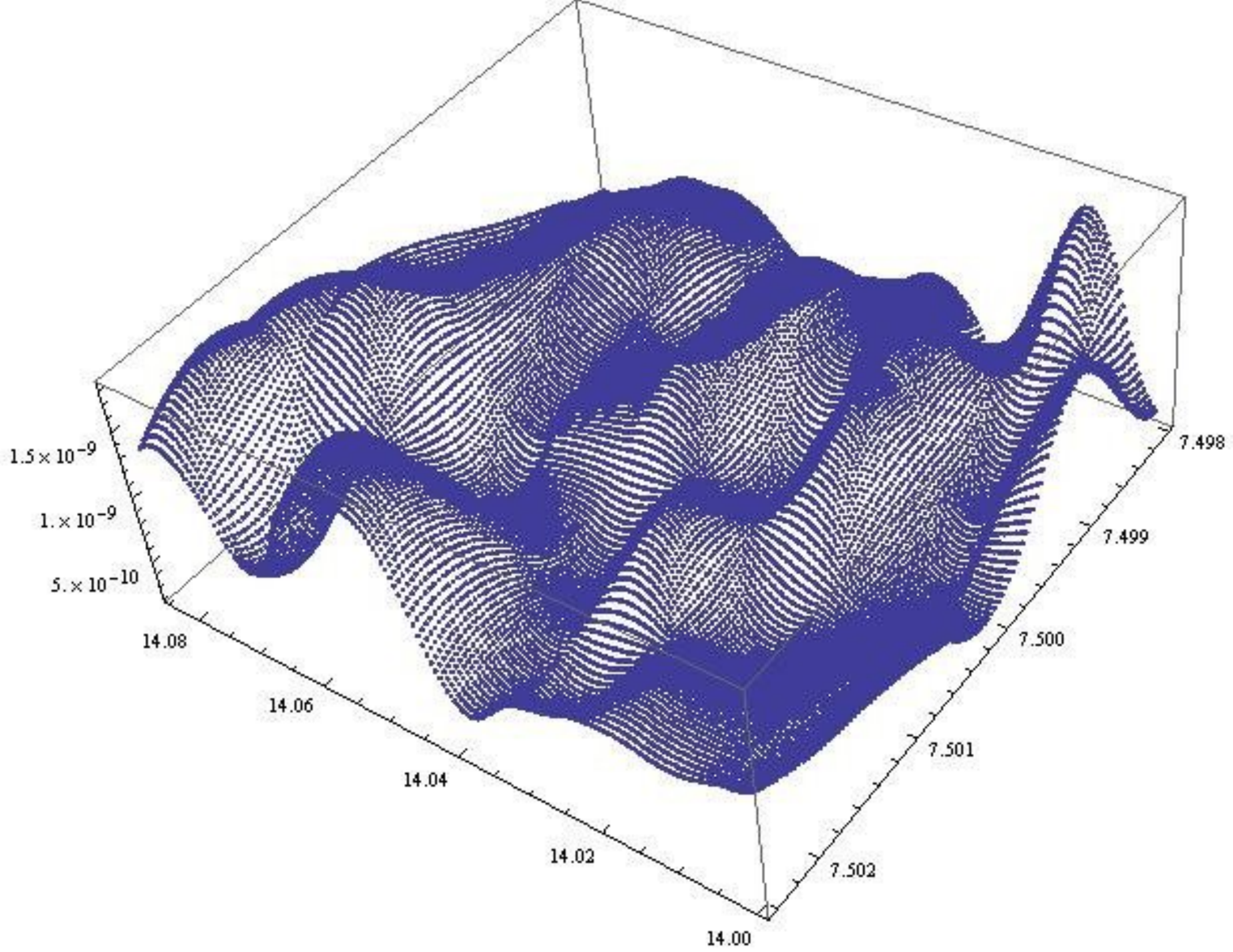}
\caption{An illustration of potentials with regularly placed bumps (left panel) and random 
bumps (right panel). The size of bumps in the actual calculations are much smaller.}
\label{fig_toypotential}
\end{figure}

\subsection{Results}

The probability of bifurcation is expressed as a two dimensional plot. The ellipticity $ \xi $ and 
$\Delta\varphi_1 / H $  are used to quantify our parameter space.

\begin{figure}[htbp]
\centering
\includegraphics[width=0.45\textwidth]{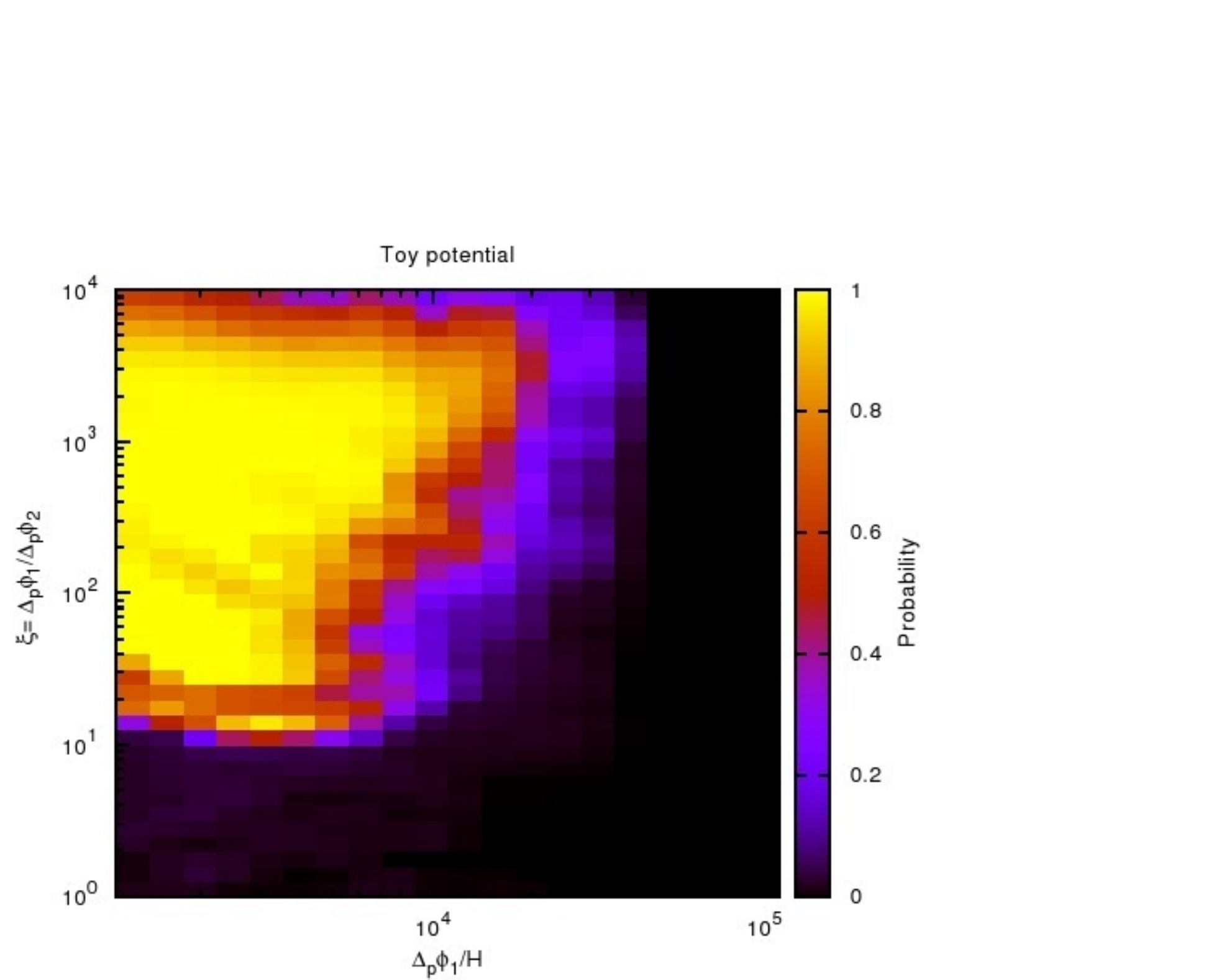}\\
\includegraphics[width=0.45\textwidth]{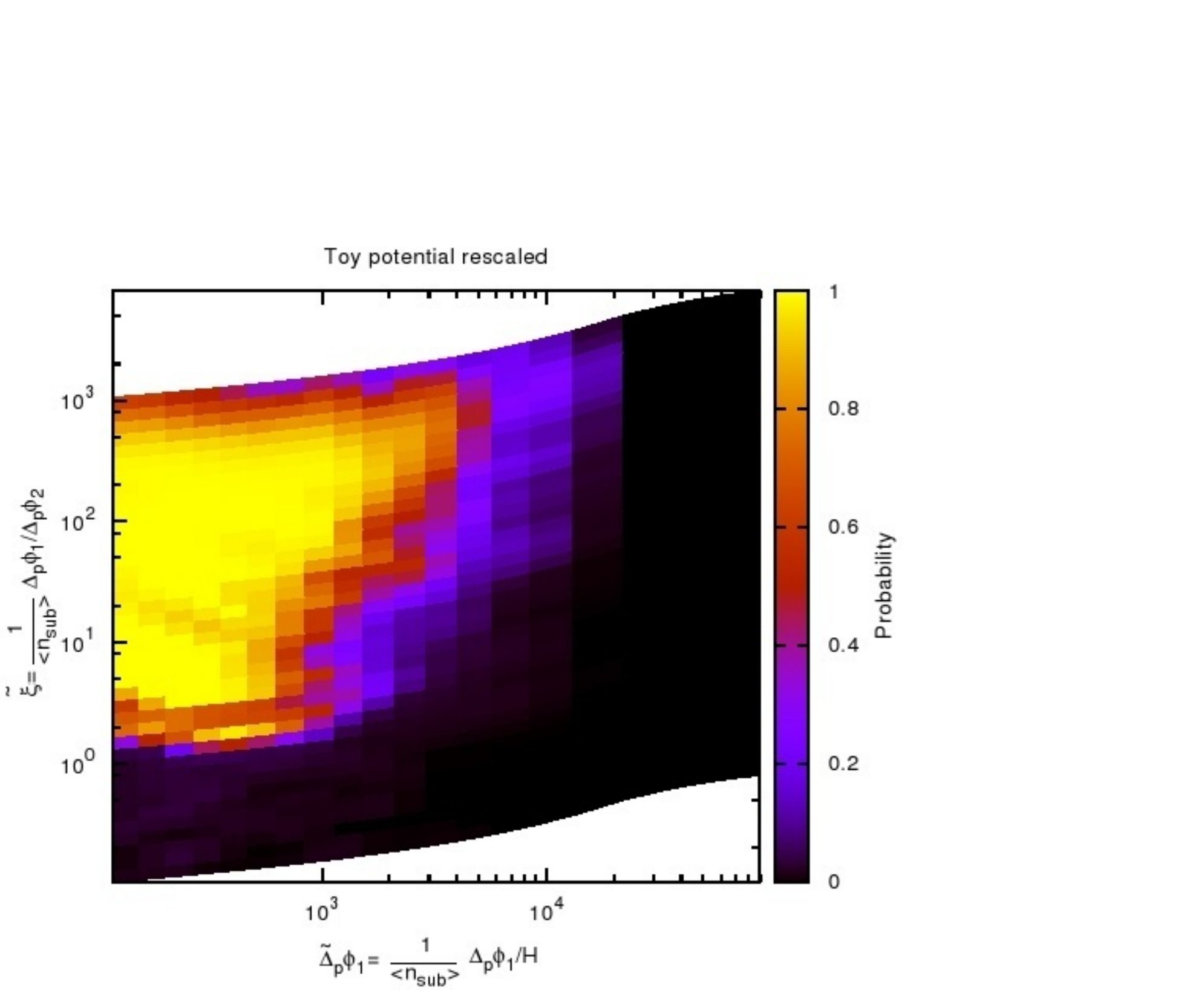}
\hspace{0.05\textwidth}
\includegraphics[width=0.45\textwidth]{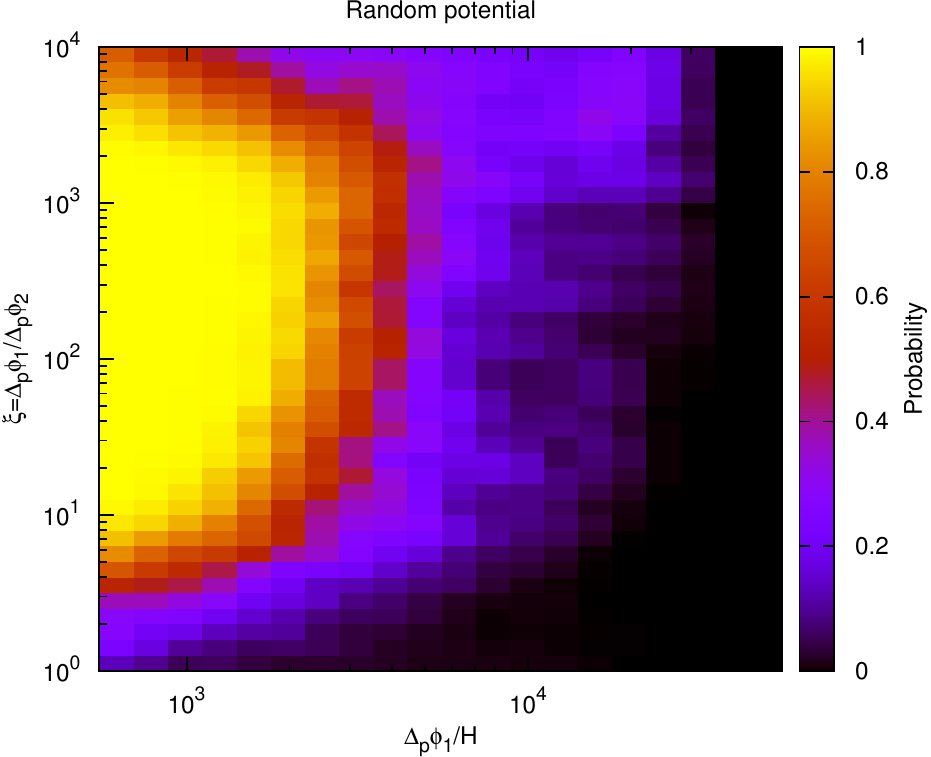}
\caption{Probability of bifurcation from numerical simulations. The upper figure is the original 
result of the toy potential with regularly spaced bumps and the bottom left panel is the same 
result taking into account the correction factor from $\langle n_\mathrm{sub} \rangle$. The 
bottom right panel is the result for a random potential.}\label{fig_heatmap}
\end{figure}

To achieve a better understanding, two types of potentials are considered: a toy potential 
with regularly arranged bumps (the left panel of figure \ref{fig_toypotential}), and a completely
random potential (the right panel of figure \ref{fig_toypotential}). The toy potential is faced with 
the same problem as the analytic model if we want to compare it to the random potential: we 
must take into account the factor $\langle n_\mathrm{sub} \rangle$ and rescale the parameter 
axes accordingly.  The latter is what we want eventually. Note that the figures are only for 
illustrative purpose, and the actual size of the bumps are much smaller. The results are 
shown in Figure \ref{fig_heatmap}. The plots can be compared with 
Figure \ref{fig_heatmap_a}, which is obtained from the analytic model.

The shape of the maps from our numerical results and based on the analytical analysis 
agree with each other when $\Delta_p\varphi_1/H < 10^4$. When 
$\Delta_p\varphi_1/H > 10^4$, then in the numerical simulations the bifurcation probability 
decreases, whereas the analytical calculation does not have lead to such a decrease.
The difference arises because in the analytical analysis we are assuming that the
trajectory encounters a large number of bumps. However,  when $\Delta_p\varphi_1/H>10^4$, 
the above assumption becomes invalid. We also note that the analytical result under-estimates 
the bifurcation probability by a factor of about 10.

\begin{figure}[htbp]
\centering
\includegraphics[width=0.45\textwidth]{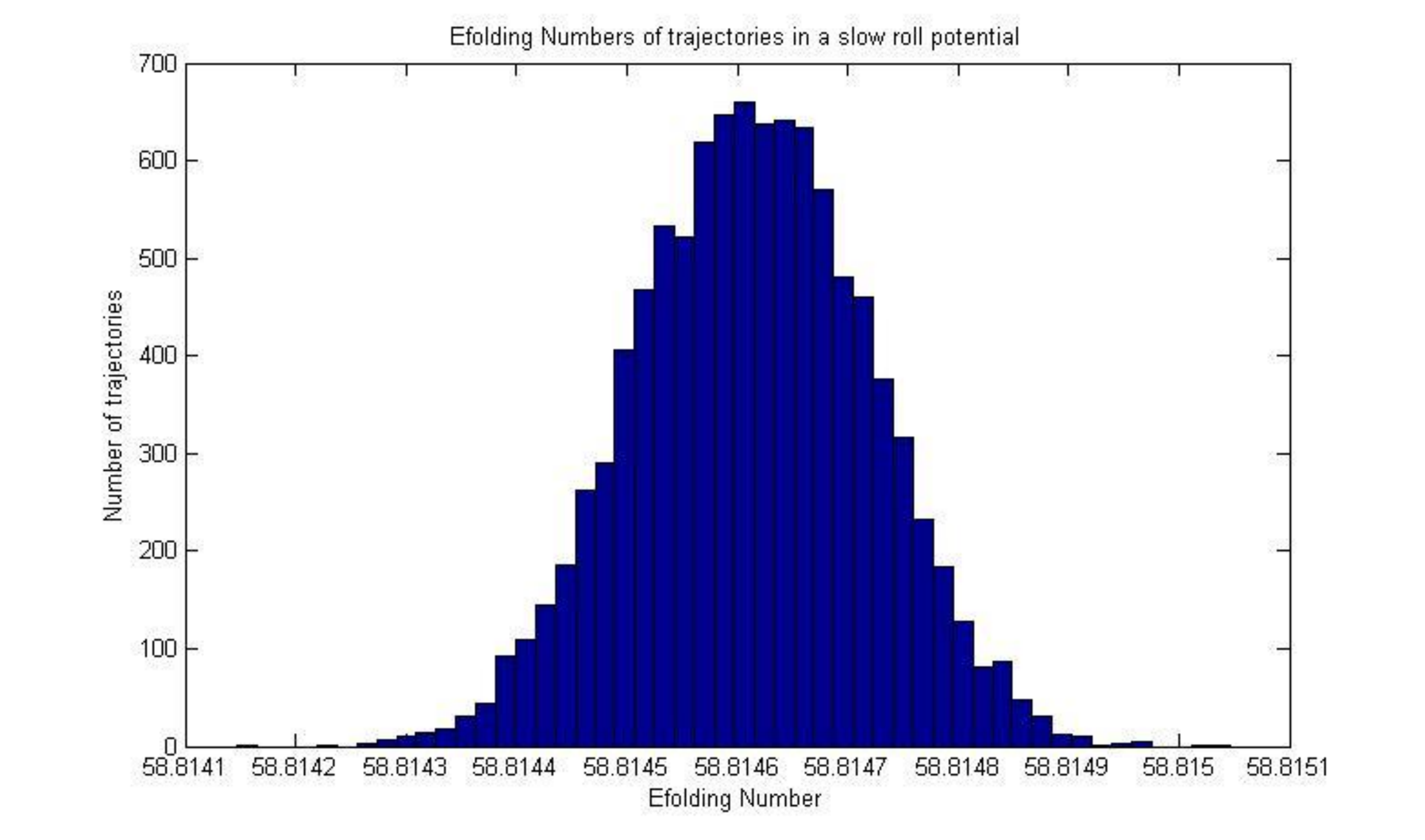}
\hspace{0.05\textwidth}
\includegraphics[width=0.45\textwidth]{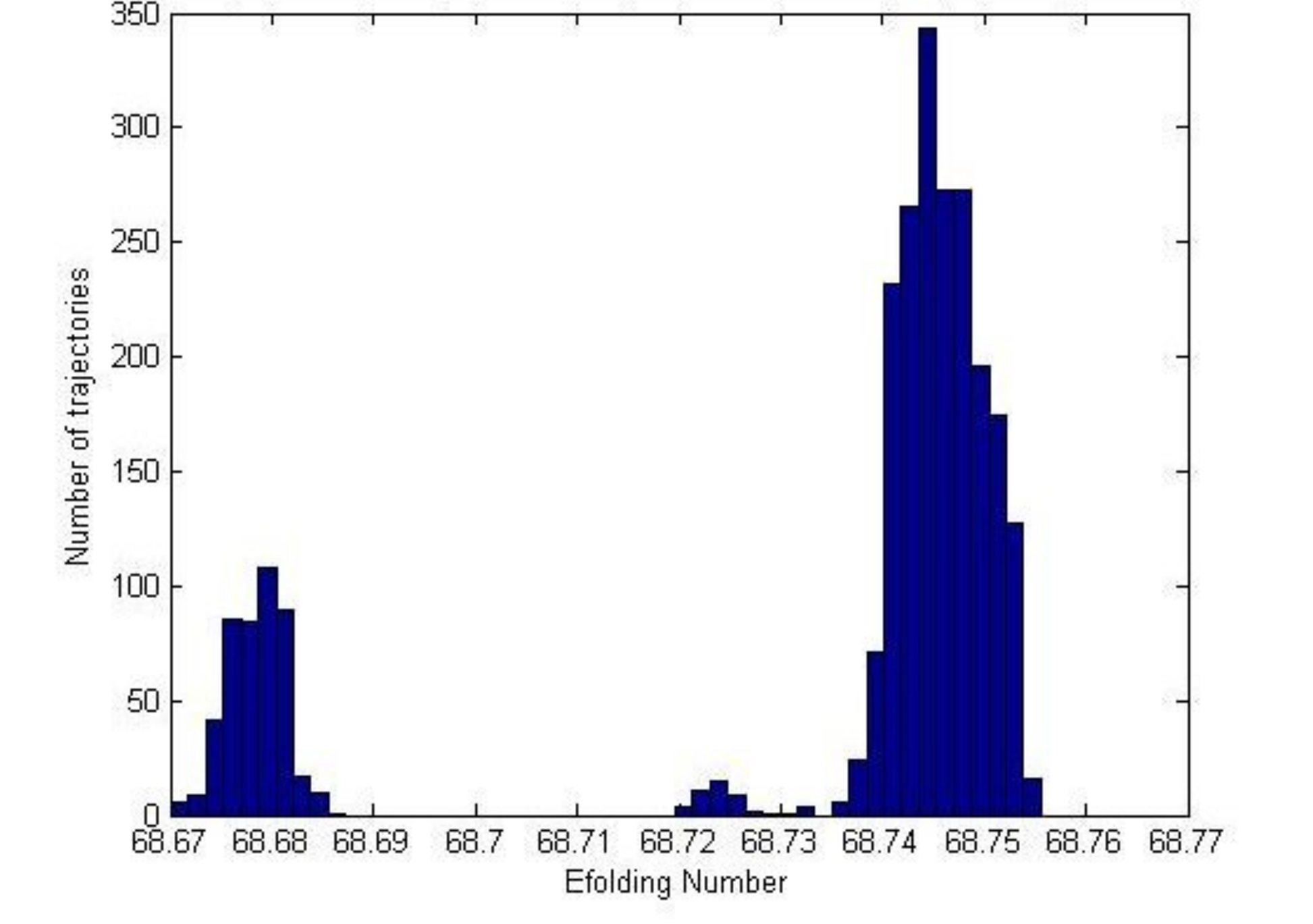}
\caption{Probability distribution for the fluctuations based on the $\delta N$ formalism. The 
left panel is for trajectories without bifurcation, and the right panel is for trajectories with a 
large probability of bifurcation, without symmetry protection or fine tuning.}\label{fig_nonG}
\end{figure}

Using the numerical code, we can also study explicitly what observational
problems occur in the case that the bifurcation probability is large. In 
Figure \ref{fig_nonG}, we show two runs without (left panel) and with (right panel) 
bifurcations. It is shown that when the bifurcation rate is large, the distribution of
e-folding numbers in local patches of the universe becomes highly non-Gaussian, 
which is not consistent with current observations.

\section{Conclusion and Discussion}

To conclude, we have considered  the probability of bifurcation of the inflaton trajectory 
in a (possibly stringy) inflationary landscape. A phase transition from bifurcating dynamics to 
absence of bifurcations is identified. The phase transition line has been found as a 
function of the mean distance of the bumps ($\Delta_p\varphi_1/H$) and of the 
ellipticity parameter $\Delta_p\varphi_1/\Delta_p\varphi_2$. 

There are a number of interesting questions in this direction which we have 
not addressed in the present work.
For example, the analytical analysis has under-estimated the bifurcation probability 
by a factor of about 10. Perhaps we are simply using too strong a criterium for bifurcation. 
But it is also possible that there are other effects leading to bifurcation, that we do not 
understand currently. Also, in the analysis leading to the results of 
Figure \ref{fig_heatmap_a} we only considered the effect from exponential growth of 
isocurvature perturbations. The random walk may also contribute in the region 
$\Delta_p\varphi_1/H < 10^3$. A combined analysis is needed to get a more 
precise analytical prediction.

As another example, as we mentioned in the introduction, a string landscape could 
have a large number of modular fields - maybe $\mathcal{O}(100)$. A lot of them 
may be light and have dynamics. We only considered two of them. It remains 
interesting to study the behavior in a many-dimensional random potential, and 
to determine how the bifurcation probability scales as a function of dimension of field space.

Also, a many-dimensional field space opens up the possibility of bifurcation in 
isocurvature directions. The bifurcation in isocurvature directions could have 
completely different predictions compared with bifurcations in the curvature direction, 
and this issue has not been investigated.

In the present work, we have used the separate universe approximation and neglected 
the field gradient terms. It would be interesting to add those terms back and do a full 
simulation in position space. Our current analysis does not consider the effect from the 
boundary regions between bubbles following different trajectories. These regions 
should become domain walls for a period of time. In the case of rare bifurcations, 
the trajectories could recombine, and thus there is no domain wall problem. 
However, it remains interesting to see whether these wall-like objects could lead 
to observationally interesting predictions, or rule out the whole model.

Finally, in the case of rare bifurcations, i.e. in the parameter regime on the boundary of the 
black regime in Figures \ref{fig_heatmap_a} and \ref{fig_heatmap}, it would be interesting to see 
what kinds of bifurcations are more natural in a landscape. The investigation of this
question may lead to a more precise correspondence between the string landscape and 
observations such as the CMB cold spot.

\section*{Acknowledgment}

This research is supported in part by an NSERC Discovery Grant, by funds from the 
CRC program, and by a Killam Research Fellowship to RB.
YW acknowledges grants from McGill University, the Institute of 
Particle Physics (Canada) and the Foundational Questions Institute.


\begin{thebibliography}{99}

\bibitem{Li:2009sp}
  M.~Li, Y.~Wang,
  ``Multi-Stream Inflation,''
  JCAP {\bf 0907}, 033 (2009).
  [arXiv:0903.2123 [hep-th]].

\bibitem{Li:2009me}
  S.~Li, Y.~Liu, Y.~-S.~Piao,
  ``Inflation in Web,''
  Phys.\ Rev.\  {\bf D80}, 123535 (2009).
  [arXiv:0906.3608 [hep-th]].

\bibitem{Wang:2010rs}
  Y.~Wang,
  ``Multi-Stream Inflation: Bifurcations and Recombinations in the
  Multiverse,''
  Journal of Cosmology, 2010, Vol 4, pages 744-759
  [arXiv:1001.0008 [hep-th]].

\bibitem{Afshordi:2010wn}
  N.~Afshordi, A.~Slosar, Y.~Wang,
  ``A Theory of a Spot,''
  [arXiv:1006.5021 [astro-ph.CO]].

\bibitem{Bousso:2000xa}
  R.~Bousso and J.~Polchinski,
  ``Quantization of four-form fluxes and dynamical neutralization of the
  cosmological constant,''
  JHEP {\bf 0006}, 006 (2000)
  [arXiv:hep-th/0004134].

\bibitem{Giddings:2001yu}
  S.~B.~Giddings, S.~Kachru and J.~Polchinski,
  ``Hierarchies from fluxes in string compactifications,''
  Phys.\ Rev.\  D {\bf 66}, 106006 (2002)
  [arXiv:hep-th/0105097].

\bibitem{Kachru:2003aw}
  S.~Kachru, R.~Kallosh, A.~Linde and S.~P.~Trivedi,
  ``De Sitter vacua in string theory,''
  Phys.\ Rev.\  D {\bf 68}, 046005 (2003)
  [arXiv:hep-th/0301240].

\bibitem{Douglas:2003um}
  M.~R.~Douglas,
  ``The statistics of string/M theory vacua,''
  JHEP {\bf 0305}, 046 (2003)
  [arXiv:hep-th/0303194].

\bibitem{Dvali}
G.~R.~Dvali and S.~H.~H.~Tye,
  ``Brane inflation,''
  Phys.\ Lett.\ B {\bf 450}, 72 (1999)
  [hep-ph/9812483].
  
\bibitem{Stephon}
S.~H.~S.~Alexander,
  ``Inflation from D - anti-D-brane annihilation,''
  Phys.\ Rev.\ D {\bf 65}, 023507 (2002)
  [hep-th/0105032].
  
\bibitem{Burgess}
C.~P.~Burgess, M.~Majumdar, D.~Nolte, F.~Quevedo, G.~Rajesh and R.~-J.~Zhang,
  ``The Inflationary brane anti-brane universe,''
  JHEP {\bf 0107}, 047 (2001)
  [hep-th/0105204].

\bibitem{stringinflationrevs}
C.~P.~Burgess and L.~McAllister,
  ``Challenges for String Cosmology,''
  Class.\ Quant.\ Grav.\  {\bf 28}, 204002 (2011)
  [arXiv:1108.2660 [hep-th]];\\
   L.~McAllister and E.~Silverstein,
  ``String Cosmology: A Review,''
  Gen.\ Rel.\ Grav.\  {\bf 40}, 565 (2008)
  [arXiv:0710.2951 [hep-th]];\\
  S.~H.~Henry Tye,
  ``Brane inflation: String theory viewed from the cosmos,''
  Lect.\ Notes Phys.\  {\bf 737}, 949 (2008)
  [arXiv:hep-th/0610221];\\
    J.~M.~Cline,
  ``Inflation from string theory,''
  hep-th/0501179;\\
  C.~P.~Burgess,
  ``Inflationary string theory?,''
  Pramana {\bf 63}, 1269 (2004)
  [hep-th/0408037].

\bibitem{Tye:2008ef} 
  S.~-H.~H.~Tye, J.~Xu and Y.~Zhang,
  ``Multi-field Inflation with a Random Potential,''
  JCAP {\bf 0904}, 018 (2009)
  [arXiv:0812.1944 [hep-th]].

\bibitem{Agarwal:2011wm} 
  N.~Agarwal, R.~Bean, L.~McAllister and G.~Xu,
  ``Universality in D-brane Inflation,''
  JCAP {\bf 1109}, 002 (2011)
  [arXiv:1103.2775 [astro-ph.CO]].

\bibitem{Frazer:2011br} 
  J.~Frazer and A.~R.~Liddle,
  ``Multi-field inflation with random potentials: field dimension, feature scale and non-Gaussianity,''
  [arXiv:1111.6646 [astro-ph.CO]].

\bibitem{Freese:2004vs}
  K.~Freese and D.~Spolyar,
  ``Chain inflation: 'Bubble bubble toil and trouble',''
  JCAP {\bf 0507}, 007 (2005)
  [arXiv:hep-ph/0412145].

\bibitem{Huang:2007ek}
  Q.~-G.~Huang,
  ``Simplified chain inflation,''
  JCAP {\bf 0705}, 009 (2007).
  [arXiv:0704.2835 [hep-th]].

\bibitem{Chialva:2008zw}
  D.~Chialva, U.~H.~Danielsson,
  ``Chain inflation revisited,''
  JCAP {\bf 0810}, 012 (2008).
  [arXiv:0804.2846 [hep-th]].

\bibitem{Ashoorioon}
A.~Ashoorioon, K.~Freese and J.~T.~Liu,
 ``Slow nucleation rates in Chain Inflation with QCD Axions or Monodromy,'
 Phys.\ Rev.\  D {\bf 79}, 067302 (2009)
 [arXiv:0810.0228 [hep-ph]].

\bibitem{Cline:2011fi}
  J.~M.~Cline, G.~D.~Moore, Y.~Wang,
  ``Chain Inflation Reconsidered,''
  [arXiv:1106.2188 [hep-th]].

\bibitem{Chung:1999ve}
  D.~J.~H.~Chung, E.~W.~Kolb, A.~Riotto, I.~I.~Tkachev,
  ``Probing Planckian physics: Resonant production of particles during inflation and features in the primordial power spectrum,''
  Phys.\ Rev.\  {\bf D62}, 043508 (2000).
  [hep-ph/9910437].

\bibitem{Romano:2008rr}
  A.~E.~Romano, M.~Sasaki,
  ``Effects of particle production during inflation,''
  Phys.\ Rev.\  {\bf D78}, 103522 (2008).
  [arXiv:0809.5142 [gr-qc]].

\bibitem{Barnaby:2009dd}
  N.~Barnaby and Z.~Huang,
  ``Particle Production During Inflation: Observational Constraints and
  Signatures,''
  Phys.\ Rev.\  D {\bf 80}, 126018 (2009)
  [arXiv:0909.0751 [astro-ph.CO]].

\bibitem{Barnaby:2010ke}
  N.~Barnaby,
  ``On Features and Nongaussianity from Inflationary Particle Production,''
  Phys.\ Rev.\  D {\bf 82}, 106009 (2010)
  [arXiv:1006.4615 [astro-ph.CO]].

\bibitem{Battefeld:2010sw}
  D.~Battefeld, T.~Battefeld,
  ``A Terminal Velocity on the Landscape: Particle Production near Extra Species Loci in Higher Dimensions,''
  JHEP {\bf 1007}, 063 (2010).
  [arXiv:1004.3551 [hep-th]].

\bibitem{Battefeld:2011yj}
  D.~Battefeld, T.~Battefeld, C.~Byrnes, D.~Langlois,
  ``Beauty is Distractive: Particle production during multifield inflation,''
  JCAP {\bf 1108}, 025 (2011).
  [arXiv:1106.1891 [astro-ph.CO]].

\bibitem{Tye:2009ff}
  S.~-H.~H.~Tye, J.~Xu,
  ``A Meandering Inflaton,''
  Phys.\ Lett.\  {\bf B683}, 326-330 (2010).
  [arXiv:0910.0849 [hep-th]].

\bibitem{Chen:2009we}
  X.~Chen, Y.~Wang,
  ``Large non-Gaussianities with Intermediate Shapes from Quasi-Single Field Inflation,''
  Phys.\ Rev.\  {\bf D81}, 063511 (2010).
  [arXiv:0909.0496 [astro-ph.CO]].

\bibitem{Chen:2009zp}
  X.~Chen, Y.~Wang,
  ``Quasi-Single Field Inflation and Non-Gaussianities,''
  JCAP {\bf 1004}, 027 (2010).
  [arXiv:0911.3380 [hep-th]].

\bibitem{TB}
J.~H.~Traschen and R.~H.~Brandenberger,
  ``Particle Production During Out-of-equilibrium Phase Transitions,''
  Phys.\ Rev.\ D {\bf 42}, 2491 (1990);\\
  Y.~Shtanov, J.~H.~Traschen and R.~H.~Brandenberger,
  ``Universe reheating after inflation,''
  Phys.\ Rev.\ D {\bf 51}, 5438 (1995)
  [hep-ph/9407247].
  
\bibitem{KLS}
L.~Kofman, A.~D.~Linde and A.~A.~Starobinsky,
  ``Towards the theory of reheating after inflation,''
  Phys.\ Rev.\ D {\bf 56}, 3258 (1997)
  [hep-ph/9704452].
  
\bibitem{Allahverdi}
R.~Allahverdi, R.~Brandenberger, F.~-Y.~Cyr-Racine and A.~Mazumdar,
  ``Reheating in Inflationary Cosmology: Theory and Applications,''
  Ann.\ Rev.\ Nucl.\ Part.\ Sci.\  {\bf 60}, 27 (2010)
  [arXiv:1001.2600 [hep-th]].
  
 \bibitem{Starob}
A.~A.~Starobinsky,
  ``Stochastic De Sitter (inflationary) Stage In The Early Universe,''
  In *De Vega, H.j. ( Ed.), Sanchez, N. ( Ed.): Field Theory, Quantum Gravity and Strings*, 107-126.

\end{thebibliography}
\end{document}